# The Hurricane Track Fit Consensus Model For Improving Hurricane Forecasting

Nathan Ginis[1] and Timothy Marchok[2]

1  Pierrepont High School, Westport, CT
2  NOAA / Geophysical Fluid Dynamics Laboratory, Princeton, NJ

**Abstract**

We present a new method for creating a model consensus to improve real-time hurricane track prediction. The method is based on the statistical fitting of historic numerical model track forecasts to the observed storm positions and learning from their historical errors and biases. Our method is closest to the Hurricane Forecast Improvement Program (HFIP) Corrected Consensus Approach (HCCA) methodology (Simon et al. 2018) while using an alternative model formulation.  Our method creates a separate consensus model for each forecast hour making it possible to independently correct the bias of each input model for that specific hour. This approach, which we call the Hurricane Track Fit (HFIT) model, is computationally efficient and scalable to additional numerical models as input, and it produces interpretable coefficients weighing model contributions.

The new method is evaluated for the 2014-2021 hurricane seasons in the Atlantic basin using the input from the best-performing operational track forecast guidance at the National Hurricane Center (NHC): the U.S. National Weather Service (NWS) Global Forecast System (GFS) deterministic (AVNI) and ensemble mean (AEMI) models, European Centre for Medium-Range Weather Forecasts (ECMWF) deterministic model (EMXI), the NWS Hurricane Weather Research and Forecasting model (HWFI) and the NHC equally weighted numerical model track consensus (TVCA). The results of the cross-validation for the 2014-2021 hurricane track dataset show that the HFIT consensus model consistently reduces the track forecast errors compared to those from the input models and the official NHC forecasts (OFCL). For example, at 24h the HFIT track forecast errors are smaller by 18.5% and 15.6% than those in AVNI and EMXI respectively, and 23% and 15% smaller at 72h. The HFIT forecasts show a reduction of errors compared to OFCL by 8.1% at 24h and 7.5% at 72h. We also discuss the successful real-time operational performance of HFIT during the 2022 hurricane season.

## 1. Introduction

The National Hurricane Center (NHC) uses a variety of multimodel consensus methods to improve the official tropical cyclone (TC) track forecasts. These methods began as simple, equally weighted averages of several models that Goerss (2000) found to consistently outperform each of the individual member models that made up the consensus.  These were improved upon by the multimodel superensemble technique (Krishnamurti et al. 1999, 2000a,b, 2001; Krishnamurti 2003) that addressed the shortcoming of the equally weighted consensus by applying unequal weights to bias-corrected member model output. These became operationalized by various forecasting centers for predicting TC tracks (e.g., Williford et al. 2003; Krishnamurti et al. 2011).

This study reports on the development and implementation of the Hurricane Track Fit (HFIT) consensus model for TC track prediction. The HFIT model is most similar to the HFIP Corrected Consensus Approach (HCCA) methodology (Simon et al. 2018), which is used for forecasting TC track and intensity. However, it uses a different consensus model formulation. Rather than implementing an "increment" approach originally developed by Krishnamurti et al. (1999) and adopted by Simon et al. (2018), the HFIT model employs a "direct fit" approach. Our method uses a separate consensus model that individually fits the predicted storm location by each input model to the observed location for each forecast hour, e.g., 12, 24, 36, 48, 72, and 120h. Developing a separate consensus model for each forecast hour makes it possible to correct the bias of each input model for that specific hour. It also opens the possibility of choosing different input models for each hour, depending on their skill in predicting those hours.

The HFIT model was developed as an alternative to the other consensus models and aims to provide forecasters with flexibility and transparency in choosing input models so that potential improvements to the technique could be easily developed, tested, and implemented.

## 2. Methodology

The HFIT model uses the forecasted change in latitude and longitude from the input models for each forecast hour and fits them to the observed change in latitude and longitude of the storm over that hour. For a particular forecast hour *h*, HFIT employs the following multilinear regression equations that are used for model fitting during the training phase.

$$LAT_{t+h} - LAT_t = a + \sum_{i=1}^{N} b_i \left( MODELLAT_{i,t+h} - LAT_t \right) \tag{1}$$

$$LON_{t+h} - LON_t = c + \sum_{i=1}^{N} d_i \left( MODELLON_{i,t+h} - LON_t \right) \tag{2}$$

Where $LAT(LON)_t$ and $LAT(LON)_{t+h}$ are the observed latitude(longitude) of the storm center at times $t$ and $t + h$, $MODELLAT(LON)_{i,t+h}$ is the forecast latitude(longitude) for hour *h* of model *i* at time $t+h$, and $N$ is the number of the input models used.

The output of the model fitting is a set of weighting coefficients $\{a, b_i.., b_N\}$ and $\{c, d_i.., d_N\}$ that describe the size of the effect of each input model to minimize the mean squared difference between the forecasted and observed latitude and longitude differences at each forecast hour.

For the forecast equations, we use the weighting coefficients obtained from training to predict the latitude $HFITLAT_{t+h}$ and longitude $HFITLON_{t+h}$ at each forecast hour $h$:

$$HFITLAT_{t+h} = LAT_t + a + \sum_{i=l}^{N} b_i \left( MODELLAT_{i,t+h} - LAT_t \right) \qquad (3)$$

$$HFITLON_{t+h} = LON_t + c + \sum_{i=l}^{N} d_i \left( MODELLON_{i,t+h} - LON_t \right) \qquad (4)$$

The reason a linear fit is appropriate can be seen in the correlation figures shown in Fig 1. The forecast changes in latitude and longitude of the storm center in the TVCA model are plotted vs the changes in the actual storm latitude and longitude, respectively. Similar relationships are found for all other input models used in this study (not shown).

Observed Differences in Latitude and Longitude Positions vs TVCA Differences 2014-2021

## 12hr

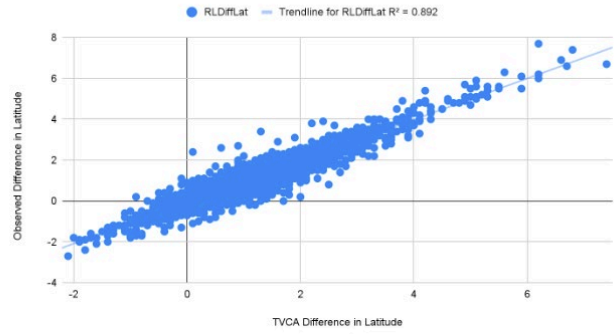

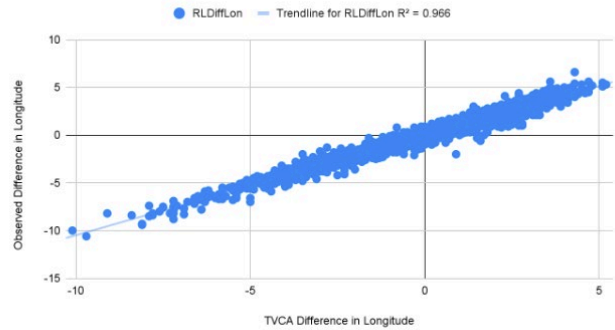

# 72hr

### Observed Difference in Latitude vs TVCA Difference in Latitude

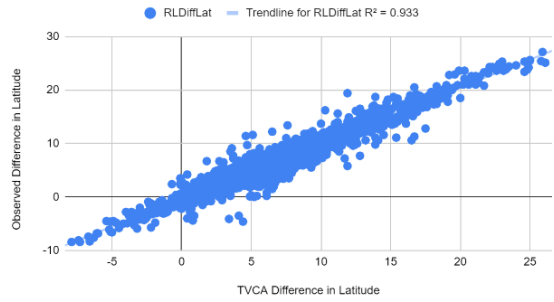

### Observed Difference in Longitude vs TVCA Difference in Longitude

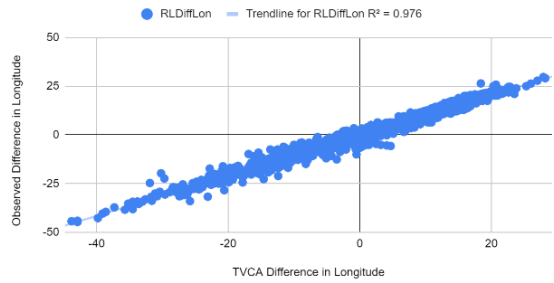

# 120hr

### Observed Difference in Latitude vs TVCA Difference in Latitude

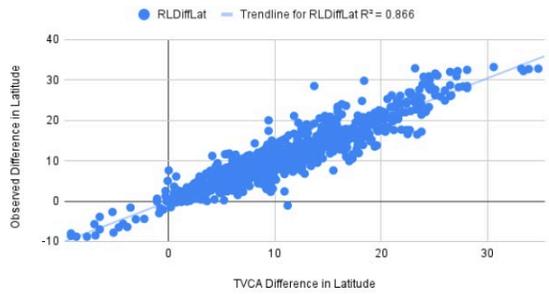

### Observed Difference in Longitude vs TVCA Difference in Longitude

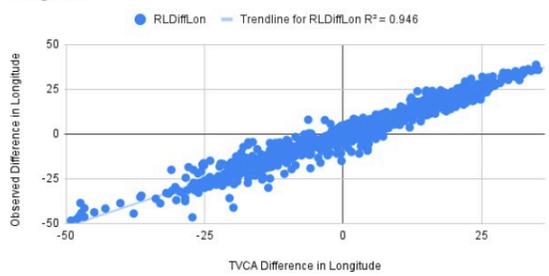

Figure 1. Correlation of latitude (top) and longitude (bottom) between the differences in the storm center positions predicted by the NHC TVCA model and the observed differences at forecast hours 12, 72, and 120 during the 2014-2021 hurricane seasons in the Atlantic Basin.

The input forecast models used in the HFIT consensus model are referred to as "interpolated," shown with model identifiers ending in "I" or "2," indicating the forecast was interpolated from the previous 6- or 12-h forecast, respectively. The model guidance at NHC is defined as either being "late" or "early," depending on its availability in a timely manner for the operational forecast cycle (see NHC 2017). Since dynamical models (regional and global) require several hours to integrate, they are generally not available to the forecasters before they issue their forecasts, 3h after the synoptic time. To reduce this timing issue, late model forecasts from the previous forecast cycle are shifted to produce what are known as the "interpolated models", so that the initial positions from the forecast issued 6 or 12h ago match the current position of the tropical cyclone.

The input models for HFIT can be selected from various dynamical and dynamical-statistical models that are available in a timely manner for real-time (operational) forecasting. HFIT currently uses five input models for hurricane track prediction. Descriptions of the HFIT input models and other hurricane datasets used in this study are given in Table 1.

Table 1. A list of the model forecast track datasets used in this study. The HFIT input models are indicated in the right-most column. Interpolation (h) refers to the operational timeliness of the models. For example, for EMXI/2, EMXI refers to the 6-h-old interpolated model, and EMX2 refers to the 12-h-old interpolated model.

| Hurricane Track Dataset | Description | Interpolation |
|---|---|---|
| AEMI | National Centers for Environmental Prediction (NCEP) Global Ensemble Forecast System (GEFS) mean | 6 |
| AVNI | NCEP GFS deterministic model | 6 |
| EMXI/2 | ECMWF deterministic model | 6/12 |
| HWFI | Hurricane Weather Research and Forecasting Model | 6 |
| TVCA | Track variable consensus (equally weighted among the top five or six track models, and includes HFIT member models AVNI, EMXI, and HWFI) | 6 |

## 3. HFIT Training and performance

The training datasets only include forecasts for lead times at which the initial and verifying classifications in the best track are either tropical depression (TD), tropical storm (TS), subtropical depression (SD), subtropical storm (SS), or hurricane (HU). At longer forecast lead times, the modeled storm becomes more likely to dissipate, thus decreasing the number of input model forecasts available for the training. Once the hurricane season ends, the storms from that season can be added to the HFIT model training dataset. Since the HFIT training process is rather efficient, it can also be done during the hurricane season by adding a tropical storm to the training set immediately after its dissipation.

### 3.1 2014-2021 hurricane seasons

The HFIT weighting coefficients for the input models were calculated during the 2014-2021 hurricane seasons using a 'cross-validation' technique, in which the entire dataset is used except for the current storm, which is withheld and then predicted. The contributions of each input model to the latitude and longitude coefficients as a function of forecast hours are presented in Fig 2. The EMXI coefficients are the largest by a significant margin across all forecast hours, within 0.5 and 1.25 for the latitude and 0.75 and 1.5 for the longitude. The coefficients for other input models, HWFI, AVNI, AEMI, and TVCA vary between -0.25 and 0.25 for the latitude and -0.5 and 0.5 for the longitude.

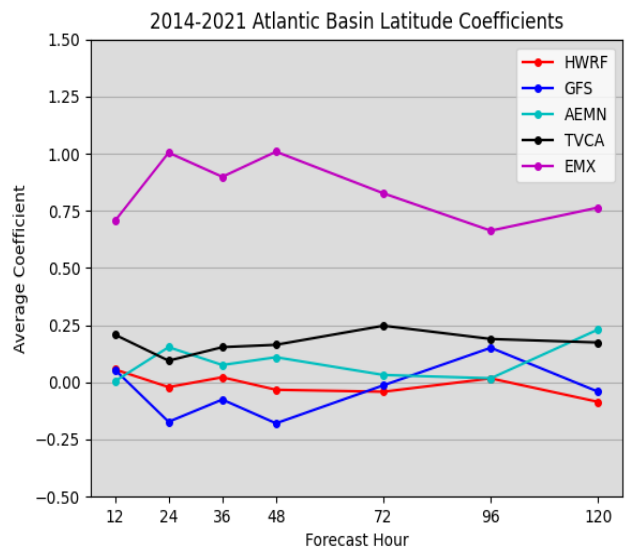

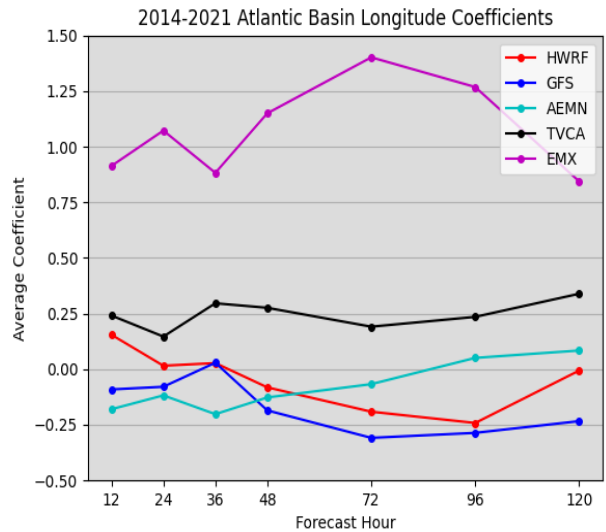

Fig. 2. Contributions of each input model to the latitude (top) and longitude (bottom) coefficients as a function of forecast hours during the 2014-2021 hurricane seasons.

The track verifications for the 2014-2021 forecasts for the HFIT model using the 'cross-validation' method in comparison to the NHC OFCL as well as the HFIT input models, HWFI, AVNI, AEMI, EMXI, and TVCA, are presented in Table 2. HFIT shows more skillful track forecasts from 12 to 72h compared to the OFCL forecasts. Beyond 72h, the difference between the HFIT model and the OFCL model was not statistically significant. The HFIT forecasts had also significantly less errors than the input models.

Table 2. Track forecast errors (n mi) for the 2014-2021 Atlantic seasons from a homogeneous sample of NHC OFCL and HFIT Forecasts, as well as the HFIT input models: HWFI, AVNI, AEMI, EMXI, and TVCA. The lowest errors at each forecast hour are shown in boldface, and N indicates the number of homogeneous verifying forecasts included in the comparison.

| Forecast Hour | N | HWFI | AVNI | AEMI | EMXI/ EMX2 | TVCA | OFCL | HFIT | % Improvement vs OFCL |
|---|---|---|---|---|---|---|---|---|---|
| 12 | 2245 | 27.9 | 26.6 | 26.8 | 25.7 | 23.8 | 24.4 | **23.2** | 4.9 |
| 24 | 2128 | 45.7 | 43.1 | 43.4 | 41.6 | 37.1 | 38.2 | **35.1** | 8.1 |
| 36 | 1998 | 63.9 | 59.9 | 60.9 | 57.6 | 51.0 | 52.1 | **47.5** | 8.8 |
| 48 | 1841 | 83.9 | 78.8 | 78.8 | 73.3 | 65.9 | 67.0 | **61.0** | 9.0 |

| 72 | 1448 | 129.8 | 119.4 | 117.1 | 108.1 | 98.4 | 99.3 | <u>**91.9**</u> | 7.5 |
| 96 | 1161 | 191.6 | 174.2 | 172.0 | 160.7 | 146.0 | 145.4 | **137.2** | 5.6 |
| 120 | 914 | 263.3 | 239.5 | 228.3 | 221.3 | 197.7 | 195.1 | **188.1** | 3.6 |

The underlined values indicate that the difference between the HFIT and OFCL is significant at the 95% level using a two-sided Student's *t* test.

To demonstrate an example of the HFIT model skill, the 5-day track forecasts for Hurricane Irma initialized at 1800 UTC 5 September 2017 are shown in Fig. 3. Table 3 demonstrates the average skill of all HFIT Hurricane Irma forecasts, and the HFIT model performed better on average than the NHC official for forecast hours 12-72.

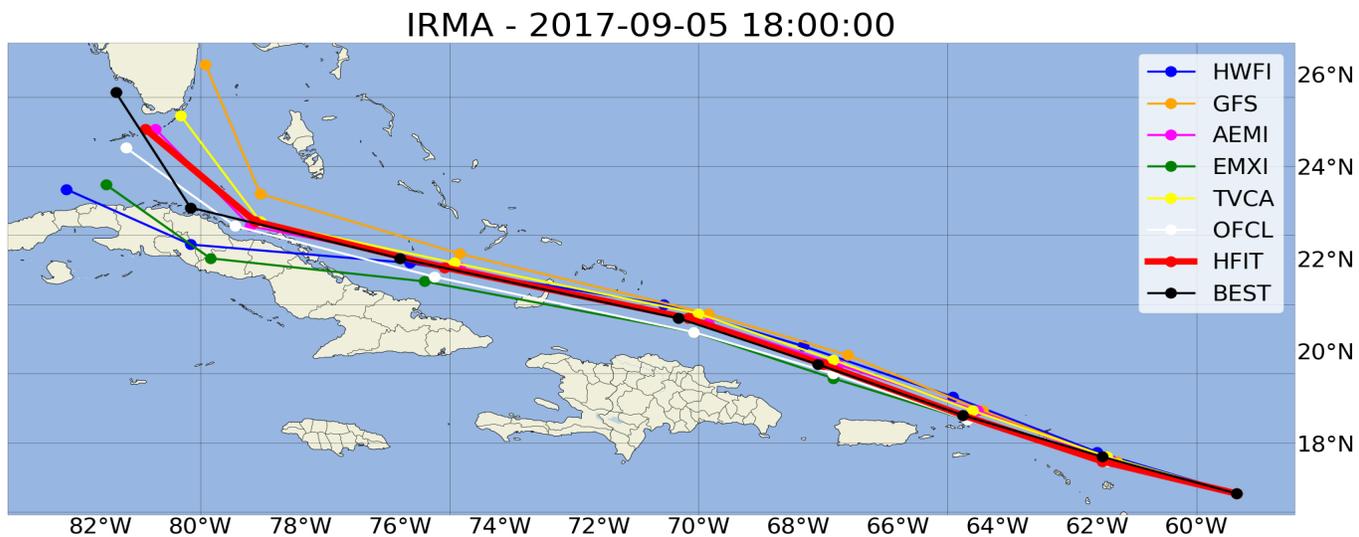

Figure 4. Track forecasts for Hurricane Irma initialized at 1800 UTC 5 September 2017.

Table 3. Mean Track Forecast Error of OFCL and HFIT during the lifecycle of Hurricane Irma (2017).  Positive % indicates improved skill.

| Forecast Hour | N | OFCL Mean Track Error (n mi) | HFIT Mean Track Error (n mi) | % Difference vs OFCL |
|---|---|---|---|---|
| 12 | 50 | 14.6 | **13.5** | 8.1 |
| 24 | 50 | 24.2 | **21.7** | 11.5 |
| 36 | 50 | 34.7 | **32.2** | 7.8 |
| 48 | 49 | 45.4 | **39.6** | 14.6 |
| 72 | 45 | 67.0 | **61.7** | 8.6 |
| 96 | **41** | **99.0** | 103.7 | -4.5 |
| 120 | **37** | **144.7** | 152.7 | -5.9 |

### 3.2 HFIT forecasts for the 2021 hurricane season

To evaluate the model skill for the entire year, HFIT was run as it would in real-time during the 2021 hurricane season with all input models available. The performance of HFIT compared to its input models and the OFCL model is shown below in Table 4.

Table 4. Track forecast errors (n mi) for the 2021 Atlantic season from a homogeneous sample of NHC OFCL and HFIT forecasts, as well as the HFIT input models: HWFI, AVNI, AEMI, EMXI, and TVCA. The lowest errors at each forecast hour are set in bold, and N indicates the number of homogeneous verifying forecasts included in the comparison.

| Forecast Hour | N | HWFI | AVNI | AEMI | EMXI/ EMX2 | TVCA | OFCL | HFIT | % Difference vs OFCL |
|---|---|---|---|---|---|---|---|---|---|
| 12 | 314 | 27.9 | 25.5 | 25.8 | 26.4 | 23.9 | 24.8 | **23.8** | 4.0 |
| 24 | 298 | 45.3 | 41.6 | 42.0 | 42.2 | 36.8 | 37.7 | **36.5** | 3.2 |
| 36 | 282 | 63.8 | 59.9 | 60.3 | 60.7 | 51.7 | 51.4 | **49.9** | 2.9 |
| 48 | 260 | 80.7 | 73.0 | 70.4 | 70.3 | 60.6 | 61.3 | **60.2** | 1.8 |
| 72 | 206 | 110.2 | 103.3 | 97.1 | 101.4 | **86.2** | 89.3 | 86.3 | 3.5 |
| 96 | 155 | 172.1 | 141.7 | 146.7 | 154.9 | 130.5 | **129.0** | 129.1 | -0.1 |

| 120 | 110 | 250.6 | 197.8 | 202.5 | 213.9 | 178.4 | **169.9** | 175.9 | -3.5 |

Through 72h, HFIT shows smaller track forecast errors averaged over the 2021 Atlantic season than the NHC official forecasts but the differences between the OFCL and the HFIT model are not significant.

### 3.3 2014-2021 hurricane seasons without EMXI/EMX2

Since the HFIT model does not currently have access to the EMXI/EMX2 model during real-time forecasting, an alternate configuration of HFIT using the other four input models (HWFI, AVNI, AEMI, and TVCA) has been developed. The contributions of each input model to the latitude and longitude coefficients as a function of forecast hours are presented in Fig 5. The TVCA coefficients are the largest by a significant margin across all forecast hours.

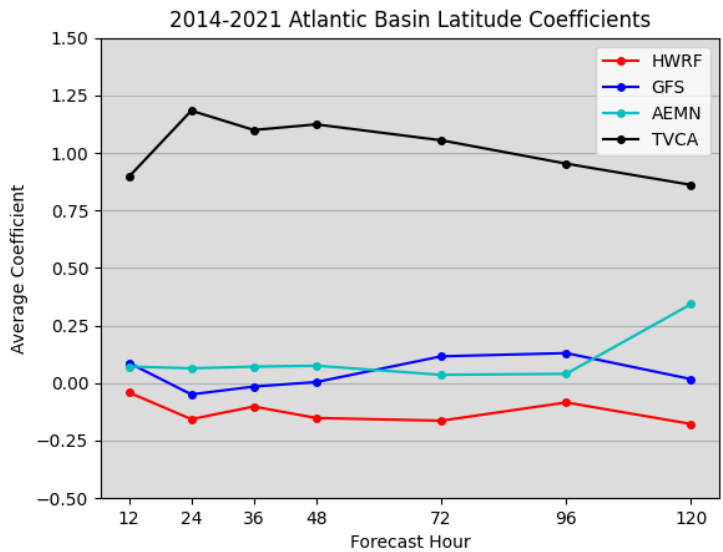

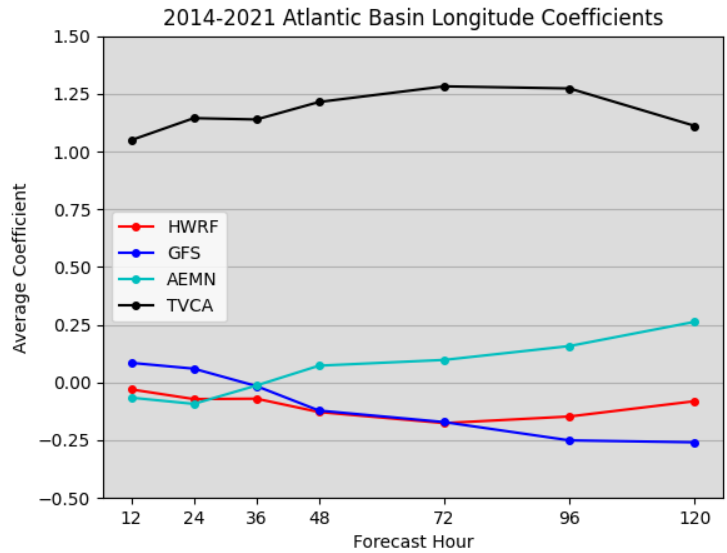

Fig. 5. Contributions of input models, HWRF, GFS, AEMN, and TVCA, to the latitude and longitude coefficients as a function of lead time during the 2014-2021 Atlantic hurricane seasons.

Table 5 indicates that at all lead times, HFIT has smaller track forecast errors than OFCL and the input models.

Table 5. Track forecast errors (n mi) for the 2014-2021 Atlantic seasons from a homogeneous sample of NHC OFCL and HFIT forecasts, as well as the HFIT input models: HWFI, AVNI, AEMI, and TVCA. The lowest errors at each forecast hour are set in boldface, and N indicates the number of homogeneous verifying forecasts included in the comparison.

| Forecast Hour | N | HWFI | AVNI | AEMI | TVCA | OFCL | HFIT | % Improvement vs OFCL |
|---|---|---|---|---|---|---|---|---|
| 12 | 2421 | 29.6 | 28.0 | 28.1 | 25.6 | 26.0 | **25.1** | 3.5 |
| 24 | 2287 | 48.7 | 45.6 | 45.5 | 39.8 | 40.9 | **<u>38.3</u>** | 6.4 |
| 36 | 2173 | 67.0 | 62.6 | 63.4 | 54.1 | 55.0 | **<u>51.6</u>** | 6.2 |
| 48 | 1997 | 86.7 | 81.3 | 81.9 | 69.0 | 70.0 | **<u>65.5</u>** | 6.4 |
| 72 | 1551 | 130.6 | 118.2 | 118.4 | 98.3 | 99.5 | **95.0** | 4.5 |
| 96 | 1266 | 187.6 | 171.5 | 171.3 | 139.5 | 142.0 | **<u>134.0</u>** | 5.6 |

| 120 | 938 | 260.4 | 239.7 | 228.7 | 191.5 | 191.3 | **184.9** | 3.3 |

The underlined values indicate that the difference between the HFIT and OFCL is significant at the 95% level using a two-sided Student's *t* test.

## 4. Real-time forecasting during the 2022 season

The HFIT (without EMXI/EMX2) model has been fully automated and provided forecasts in real-time during the 2022 hurricane season. It performed very well during Hurricane Ian. Examples of the forecasts initialized at 0600 UTC 26 Sep, 0600 UTC 27 Sep, and 0600 UTC 28 Sep are shown in Fig.6.

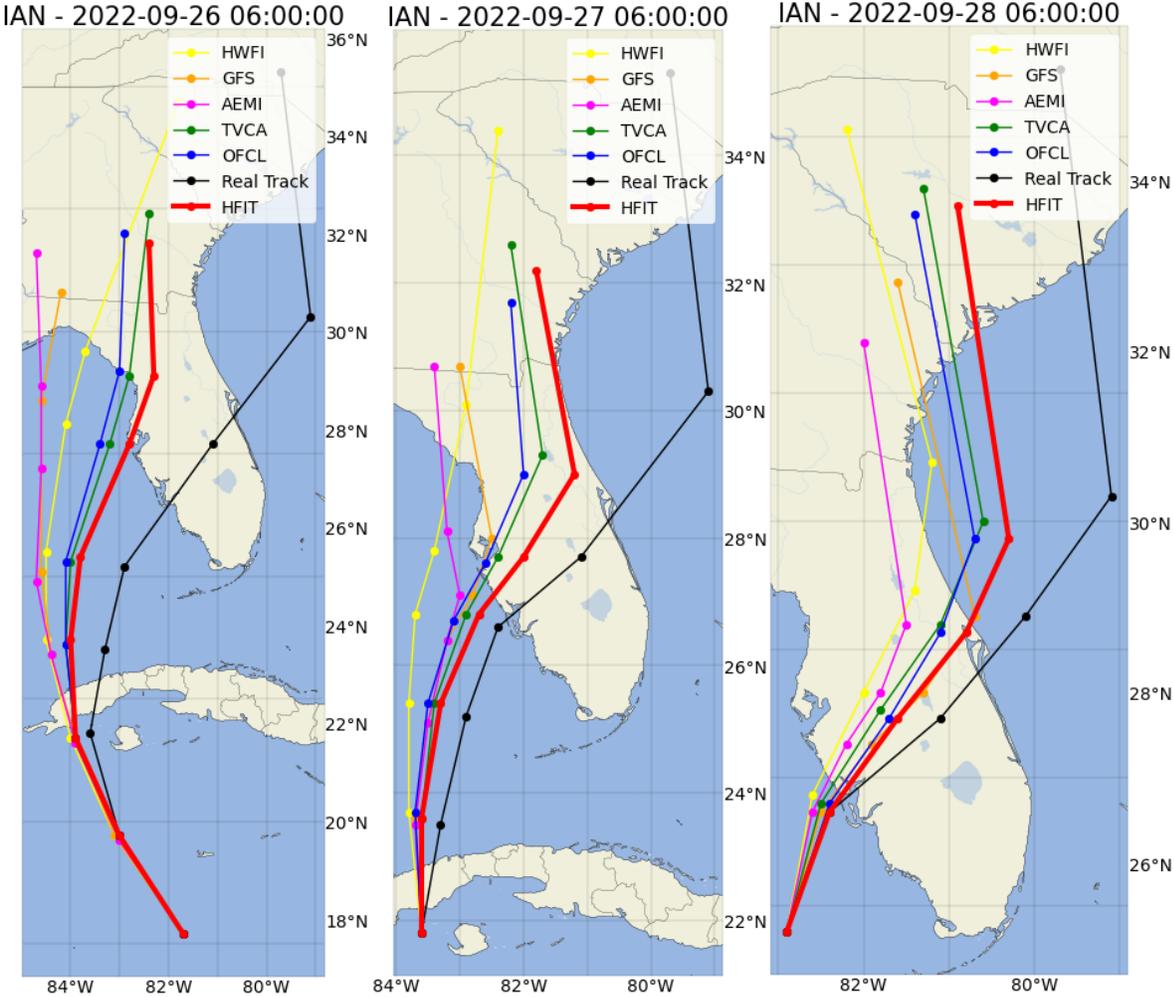

Fig 6. Hurricane Ian forecasts initialized at 0600 UTC 26 Sep, 0600 UTC 27 Sep, and 0600 UTC 28 Sep

Table 6 indicates that HFIT (without EMXI/EMX2) performed better than all input models at all lead times for Ian as well as for all lead times for the NHC Official except at 120h. For example, at a forecast lead time of 48 hours, the HFIT (without EMXI/EMX2) model performed 17.7% better than the OFCL model on average.

Table 6. Track forecast errors (n mi) for Hurricane Ian from a homogeneous sample of NHC OFCL and HFIT (without EMXI/EMX2) forecasts, as well as the HFIT (without EMXI/EMX2) input models: HWFI, AVNI, AEMI, and TVCA. The lowest errors at each forecast hour are set in bold, and N indicates the number of homogeneous forecasts included in the comparison.

| Forecast Hour | N | HWFI | AVNI | AEMI | TVCA | OFCL | HFIT | % Improvement vs OFCL |
|---|---|---|---|---|---|---|---|---|
| 12 | 31 | 23.7 | 18.6 | 22.0 | 19.2 | 20.0 | **18.1** | 9.2 |
| 24 | 31 | 43.9 | 33.5 | 40.2 | 32.1 | 33.3 | **28.2** | 15.4 |
| 36 | 29 | 70.2 | 52.8 | 64.0 | 49.5 | 50.8 | **42.5** | 16.2 |
| 48 | 27 | 96.9 | 79.4 | 91.4 | 66.7 | 69.9 | **57.5** | 17.7 |
| 72 | 23 | 148.1 | 149.3 | 158.3 | 107.6 | 112.9 | **98.5** | 12.7 |
| 96 | 20 | 204.6 | 230.1 | 218.2 | 156.4 | 159.2 | **145.9** | 8.4 |
| 120 | 16 | 233.6 | 270.0 | 241.9 | 175.3 | **168.3** | 170.1 | -1.1 |

**Summary**

This study describes the development and implementation of the Hurricane Track Fit (HFIT) consensus model for tropical cyclone track forecasts. The HFIT model uses the forecasts of several input models for track prediction in a manner similar to other corrected consensus guidance products. Unlike the previously used "increment" approach, the HFIT model employs a "direct fit" approach, such that a separate model is individually fitted to the observed track location for each track forecast hour, e.g., 12, 24, 36, 48, 72, and 120h. Unequal weighting coefficients are derived for each input model using multiple linear regression based on training forecasts extending through the 2014-2021 hurricane seasons, with separate weighting coefficients derived for each forecast hour with no influence from other forecast hours. Fitting a separate model for each hour makes it possible to calculate the best bias correction for each input model for that specific hour. It also opens the possibility of choosing different input models for each hour, depending on their prediction skill at those hours.

Conducting the cross-validation on the 2014-2021 Atlantic seasons revealed that HFIT provides skillful guidance for track forecasting. Compared to the HFIT input models, the equally weighted variable consensus, and the official NHC forecasts, HFIT had the smallest track forecast errors in the Atlantic basin from 12 to 96 h.

The HFIT model is fully automated and has produced forecasts in real-time during the 2022 hurricane season. The real-time forecasts are provided at the website: hurricaneforecast.org. The website also includes the cross-validation forecasts for 2014-2021 and an evaluation of the HFIT model performance.